# Regular Arrays of 2 nm Metal Nanoparticles for Deterministic Synthesis of Nanomaterials

Ali Javey and Hongjie Dai*

*Department of Chemistry, Stanford University, Stanford, CA 94305*
**RECEIVED DATE (automatically inserted by publisher)**; E-mail: hdai@stanford.edu

Nanoparticles with controllable sizes down to molecular dimensions have been the focus of a wide range of studies owing to their interesting physical and chemical properties and broad applications in catalysis, surface science and materials science.[1-5] For various applications, not only it is important to precisely control their size, it is also desirable to control the positioning of individual nanoparticles on substrates.[3-6] While self-assembly methods afford small nanoparticles with packed or connected structures, the formation and assembly of discrete single particles at specific sites with desired pitch require lithographic patterning techniques. Patterning of discrete nanoparticles well below 10 nm is challenging, but is highly desirable for applications such as catalytic synthesis of single-walled carbon nanotubes (SWNT) or ultra-small semiconducting nanowires from well controlled surface sites.[3,8] To our knowledge, discrete single-particle patterning for SWNT growth has not been achieved thus far.

Here, we present a novel method for patterning individual metallic nano-clusters with tunable and monodisperse sizes down to the 2 nm-scale into desired arrays, using an electron beam lithography technique that has a resolution of ~ 20nm. Furthermore, we demonstrate chemical vapor deposition (CVD) synthesis of SWNTs from regular arrays of individual ~1.5 nm particles with majority of the nanoparticles producing a SWNT at well defined locations. This reaches one of the ultimate goals of SWNT synthesis on surfaces.

Our approach involves EBL patterning and utilizes the diffusion and clustering of metal atoms on substrates at elevated temperatures (Fig.1). Wells with radius $r$~20 nm (Fig. 2A) are

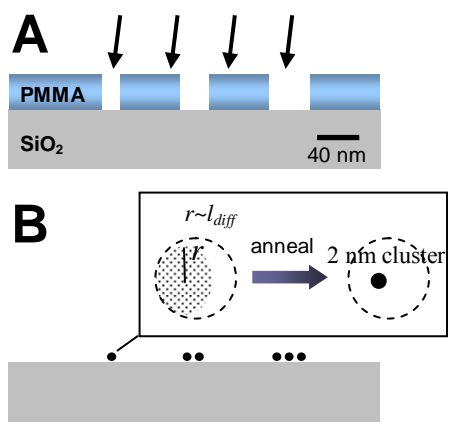

**Figure 1**. Formation of regular arrays of metal clusters down to 2nm in diameter. A) Arrays of 20 to 50 nm wells are patterned in ~100 nm thick PMMA on Si/SiO$_2$ (10 nm) substrates by electron beam lithography. Thin metal films (0.2 to 2 nm) are then deposited by an electron-beam evaporator at a 5 to 10º angle with respect to the substrate normal. B) After lift-off and annealing, one or multiple particles per well are formed depending on the well size. Inset: formation of a ~2nm particle when radius of metal-deposited area (dotted region) $r$ is ≤ metal atom diffusion length $l_{diff}$. Angle evaporation reduces $r$ by shadowing of the vertical PMMA wall. Dashed circles: bottom of the PMMA wells.

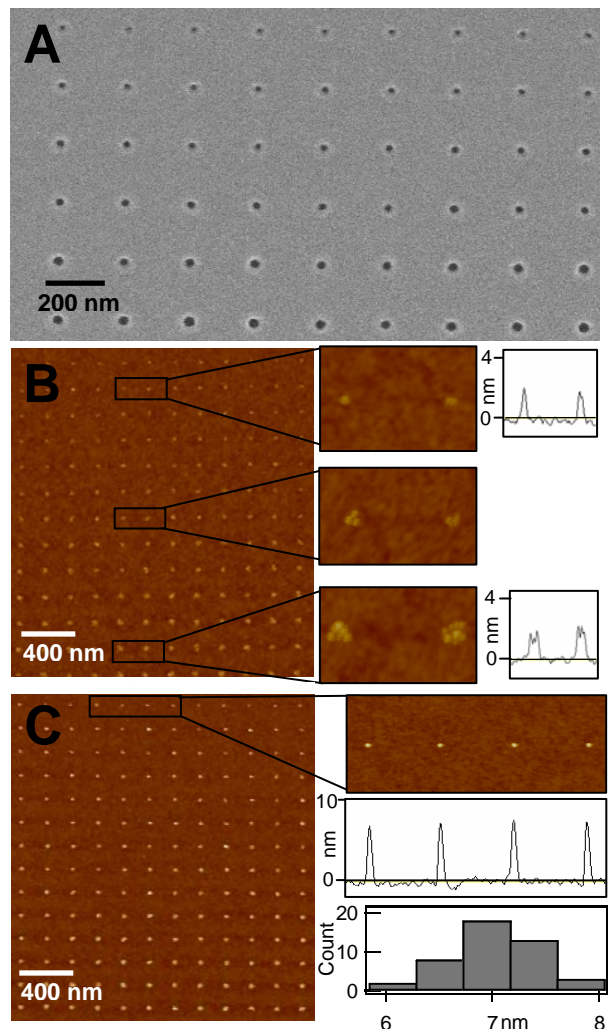

**Figure 2**. Patterning of monodispersed ~2 and 7 nm Co nanoclusters. A) Scanning electron microscopy image of patterned wells (holes) in a PMMA coated Si/SiO$_2$ substrate. Diameter of wells in each row increases by ~5% and ranges from ~20 nm on top to ~30 nm at the bottom. B) AFM image of rows of single (top right) and multiple (bottom right) ~2.2 nm Co particles and topographic line scans. The particle diameter is on the order of the measured height of ~2.2nm as the apparent width is due to AFM tip effect. C) AFM image of rows of single (top right, topography in middle; histogram at bottom) and multiple ~7 nm Co particles.

first patterned by EBL on a PMMA (100nm thick) coated SiO$_2$ substrate. Thin films (nominally 2-20 Å) of metal (Co, Fe, Pt, etc) are then evaporated at an angle (5-10º) with respect to the substrate normal. After lift-off of PMMA and thermal annealing (700-900ºC), discrete clusters are formed in an arrayed fashion (Fig.1B,2). The number of particles per site is dependent on the size of the patterned PMMA wells and the angle of evaporation

(Figure 1), and the diameter of the particles is controlled by the amount of metal deposited (by thickness monitoring). This simple approach affords arrays of various metal clusters with tunable diameters from tens of nanometers down to ~1-2 nm (Fig.2,3). Figure 2B&C show atomic force microscopy (AFM) images of arrays of ~2.2 nm and 7 nm cobalt (Co) clusters derived from ~5 and 15 Å evaporated films of Co respectively. The particles are monodispersed with a diameter distribution of ~17% and ~5% for ~2.2 and 7 nm particles respectively (for sites with one particle). This narrow distribution is among the best for nanoparticles formed in this size range by various methods.

A key to the formation of ~2nm clusters in ~20nm wells is the finite mobility and diffusion of metal atoms on $SiO_2$ at high temperatures, during which the deposited (originally dispersed) atoms irreversibly "hit and stick" to each other via metal-metal interactions and eventually form stationary clusters[1,9] (Fig.1B). By limiting the number of atoms deposited in each well (~500 atoms per well), one affords small clusters down to ~ 2nm. The mean diffusion distance $l_{diff}$ can be approximated by the Einstein relation, $l_{diff} \propto D \propto exp(-E_{diff}/2K_BT)$ where $D$ is the diffusion coefficient and $E_{diff}$ is the diffusion activation energy from site to site.[1,7] For any given metal, clustering of atoms takes place within a radius of $r \sim l_{diff}$. High temperatures enhance diffusion and $l_{diff}$, and thus cluster formation.

To obtain individual nanoparticles within a well, the size of the well should be $\leq l_{diff}$ (temperature dependent). We are able to achieve this by optimizing the annealing temperature for each metal material. The evaporation angle is used to afford better control over the number of metal atoms deposited into each well and effective reduction of the well size by 'shadowing masking'[10] using the vertical side of the PMMA wells (Fig.1A). For wells $\geq l_{diff}$, multiple clusters are formed in each well due to the finite mobility of the metal atoms and clusters, as seen in Figure 2B&C.

For Co nanoparticles, we find that the optimal annealing temperature for single particle formation in ~ 20nm wells is ~825 ºC. Multiple particles are formed in wells > 25 nm. The optimal annealing temperatures for individual ≤ 5nm Fe and Pt nanoparticles (see Fig.3 for 4-5nm Pt particles) are ~ 775 ºC and 900 ºC respectively. Lower annealing temperatures only produce multiple smaller particles per well due to incomplete aggregation. If the temperature is too high, reduction of particle size or particle disappearance is observed due to the evaporation of atoms.

Our patterning method represents a breakthrough in forming and positioning various metal clusters with highly tunable and yet monodispersed diameters down to ~1nm. The particle size distribution is partly due to the variations in PMMA-well sizes (see hole sizes in each row of Fig. 2a) resulted from fluctuations in the electron-beam flux during the exposure. Among various potential utilities, here, we demonstrate the use of this nanofabrication method to enable the ultimate patterned growth[8] of single-walled carbon nanotubes from regular arrays of ~1.5 nm Co or Fe catalytic seeds by CVD. By optimization of the growth conditions, we are able to achieve a high yield, with majority of the clusters capable of producing a SWNT (one-to-one growth) (Figure 4). The SWNT growth is enabled at the individual-particle level, and the growth locations on substrates are precisely controlled for every nanotube (at a pitch down to ~200nm thus far). The catalytic synthesis of carbon nanotubes is conducted by CVD at 825 ºC for 5 min with 300 sccm of $Ar/H_2$ (3% $H_2$) passed through an ethanol source[11] (maintained at -10 ºC). Similar one-to-one growth from 1.5 nm Fe nanoparticle arrays is also achieved by plasma assisted CVD of methane at 700 ºC.[12] Importantly, owing to the similar catalyst particles in the array, the grown SWNTs are monodisperse in diameter (Fig. 4B). This meets one of the critical challenges for nanotube electronics applications since the bandgap is governed by the tube diameter.

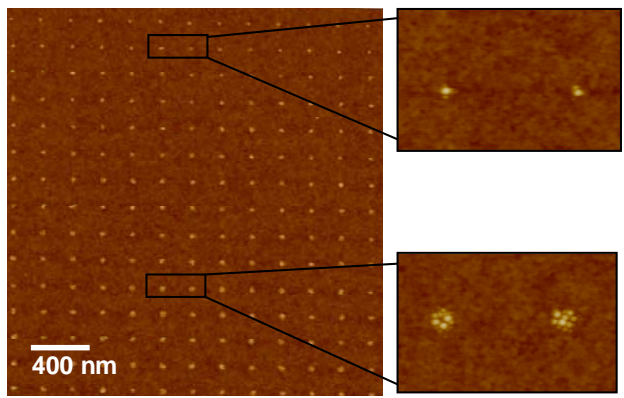

*Figure 3*. Patterned arrays of of Pt nanoparticles. Top right: a row of ~4.5 nm Pt nanoparticles formed in 20nm wells. Bottom right: a row of multiple Pt dots in bigger wells. ~1 nm Pt film was evaporated and thermal annealing was at 900 ºC in $H_2$ for 20 min.

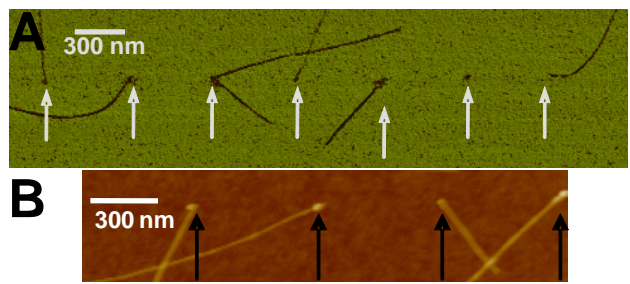

*Figure 4.* Deterministic growth of SWNTs from patterned discrete Co nanoparticles. A) & B): AFM images of nanotubes grown from arrays of 1-2 nm Co particles (pointed by arrows) with nearly one-particle to one-nanotube correspondence. The four SWNTs in B) have a mean diameter of ~1.7nm. The nanotube diameter is known to be determined by the diameter of the catalyst dots.

Besides catalytic nano-synthesis, our method may prove powerful in various scientific and technological applications ranging from catalysis, surface science, single-molecule spectroscopy, and data storage. The smallest pitch between single clusters is ~75 nm thus far, limited by the lithography system used. Potentially, the pitch can be reduced to <50nm and the throughput can be increased to full wafer-scale by patterning PMMA wells using nanoimprint lithography.[13-15]

**Acknowledgement.** This work was supported by Intel and MARCO MSD Focus Center.

Table of Contents artwork

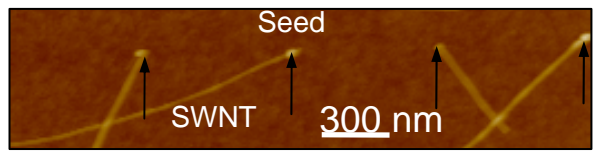

ABSTRACT FOR WEB PUBLICATION

A novel method is developed to enable the formation, positioning and patterning of individual metal nano-clusters with controllable and monodispersed sizes down to 1-2 nm-scale. The method is generic for fabricating designed arrays of virtually any type of metal nanoparticles well below 10 nm. Among wide range of potential applications in surface and materials science, the nanoparticle arrays are used for deterministic synthesis of monodispersed single-walled carbon nanotubes at individually controlled locations with near one-to-one yield, reaching one of the ultimate goals of nanotube synthesis.